\documentclass[preprint,11pt]{article}

\usepackage{arxiv}

\usepackage{amssymb}
\usepackage{float}
\usepackage{geometry}
\usepackage[caption = false]{subfig}
\usepackage{graphicx}
\usepackage{amsmath}
\usepackage{xcolor}
\usepackage{lineno}
\usepackage{multirow}
\usepackage{amsthm}
\usepackage{amsmath}
\usepackage{esvect}
\usepackage{hyperref}

\title{gPC-based robustness analysis of neural systems through probabilistic recurrence metrics}

\date{} 					

\author{Uros Sutulovic \\
	Department of Industrial Engineering\\
	University of Trento\\
	Trento, Italy \\
	\texttt{uros.sutulovic@unitn.it} \\
\And
	Daniele Proverbio \\
	Department of Industrial Engineering\\
	University of Trento\\
	Trento, Italy \\
	\texttt{daniele.proverbio@unitn.it} \\
\And
	Rami Katz \\
	School of Electrical and Computer  Engineering\\
	Tel Aviv University\\
	Tel Aviv-Yaffo, Israel \\
	\texttt{ramkatsee@tauex.tau.ac.il} \\
\And
	Giulia Giordano \\
	Department of Industrial Engineering\\
	University of Trento\\
	Trento, Italy \\
	\texttt{giulia.giordano@unitn.it} \\
}

\renewcommand{\headeright}{}
\renewcommand{\undertitle}{}

\begin{document}

\maketitle

\begin{abstract}
Neuronal systems often preserve their characteristic functions and signalling patterns, also referred to as \textit{regimes}, despite parametric uncertainties and variations.
For neural models having uncertain parameters with a known probability distribution, probabilistic robustness analysis (PRA) allows us to understand and quantify under which uncertainty conditions a regime is preserved in expectation. We introduce a new  computational framework for the efficient and systematic PRA of dynamical systems in neuroscience and we show its efficacy in analysing well-known neural models that exhibit multiple dynamical regimes: the Hindmarsh-Rose model for single neurons and the Jansen-Rit model for cortical columns. Given a model subject to parametric uncertainty, we employ generalised polynomial chaos to derive mean neural activity signals, which are then used to assess the amount of parametric uncertainty that the system can withstand while preserving the current regime, thereby quantifying the regime's robustness to such uncertainty. To assess persistence of regimes, we propose new metrics, which we apply to recurrence plots obtained from the mean neural activity signals. The overall result is a novel, general computational methodology that combines recurrence plot analysis and systematic persistence analysis to assess how much the uncertain model parameters can vary, with respect to their nominal value, while preserving the nominal regimes in expectation.
We summarise the PRA results through probabilistic regime preservation (PRP) plots, which capture the effect of parametric uncertainties on the robustness of dynamical regimes in the considered models.
\end{abstract}


\section{Introduction}
\label{sec:introduction}

In systems biology, robustness refers to the ability of a system to preserve a qualitative property despite uncertainties, variability or environmental fluctuations \cite{kitano2004biological, Khammash2016, BG2021structural, PKG2025}. In particular, systems in neuroscience can robustly preserve complex signalling patterns, often called \textit{regimes}, both at the single neuron level and at the neural population level \cite{bray2024adaptive,mirzakhalili2017synaptic,li2016robust,aerts2016brain,maclean2003activity}, despite being subject to environmental disturbances and intrinsic uncertainties \cite{koch2004biophysics,faisal2008noise}.
Understanding the mechanisms that guarantee such robustness, as well as quantifying the uncertainty levels that are compatible with preservation of a given life-sustaining signalling pattern, enables a better interpretation of recorded signals and a deeper insight into neural dynamics and neural diseases.
The study of dynamical regimes and of their robustness in neural models also supports the robust design of artificial systems inspired by dynamics in neuroscience \cite{hassabis2017neuroscience}, as well as future biomedical applications and medical interventions, \textit{e.g.}, to monitor depth of anaesthesia \cite{kuhlmann2016neural}, treat Parkinson's disease \cite{kerr2013cortical} and epilepsy \cite{subramaniyam2024sensitivity}, and contrast epileptic seizures \cite{ashourvan2020model}.

The analysis of interpretable dynamical models of neural activity is crucial to understand neural behaviours \cite{izhikevich2007dynamical}.
To assess how varying key parameters, or combinations thereof, affects the behaviour of neural systems and the resulting (number and type of) observed regimes (such as equilibria), existing works usually employ deterministic methods, such as sensitivity analysis, parameter sweeping or bifurcation analysis \cite{barrio2011parameter, ashwin2016mathematical}, where parameter values are assumed to be constant and deterministic.
These methods to assess the number and properties of dynamical regimes in a parameter space, and the regime changes associated with parameter variations, do not account for the possibility that biological parameters are affected by probabilistic uncertainties.
Also, given experimental data, traditional modelling typically employs deterministic methods to identify the single best-fit value of the system parameters that matches the observed biological phenomenon and then uses the resulting model for prediction; the usefulness and reliability of the obtained predictions critically depend on the accuracy, precision and completeness of measurements \cite{gutenkunst2007universally}.
However, parametric uncertainty and variability strongly affect real-life neuronal systems, and thus the ensuing measurements, as observed in experimental settings \cite{goaillard2021ion}.
Probabilistic parametric uncertainty causes the system parameters to take values according to a probability distribution, within feasible bounds including a nominal value, thus yielding different system realisations (such as those observed in distinct neurons within a population) that may exhibit the same qualitative physical behaviour despite having different system parameters.
This exactly matches real-life scenarios: different cells, neurons and structures with the same function do not typically share the same parameter values, since these are associated with physiological processes, such as ion fluxes through membrane channels, that are subject to variability and thus induce heterogeneity, across cells or over time \cite{nusser2009variability, goaillard2021ion}. Uncertainty and variability, therefore, are intrinsic and cannot be neglected in the study of neuronal systems.

Mathematically, uncertainty can be characterised either in a  \textit{deterministic} or in a \textit{probabilistic} setting.
Studying the effect of \emph{deterministic} parametric uncertainties is the goal of classic robustness analysis, which has a long tradition in control theory \cite{barmish1994new}.
Robustness analysis is agnostic about the exact parameter values realised in practice, and provides a “yes” or “no” answer to the question whether a property of interest (in our case, a regime) is preserved robustly, \textit{i.e.}, for all possible parameter realisations within a known bounding set in the parameter space.
The resulting sufficient conditions for the desired system performance are often overly conservative \cite[Chapter 14]{barmish1994new} and have an “all-or-nothing” nature: if, even for a single parameter choice within the bounding set, the property does not hold, then robustness cannot be guaranteed.
In nature, however, certain proportions of defective cells and processes may be present while proper biological functions are maintained nonetheless \cite{mojtahedi2016cell}. Hence, preservation of a life-sustaining behaviour “on average” (\textit{e.g.}, through a critical mass of properly functioning cells) is what truly matters. 
This observation fosters the need to investigate \emph{probabilistic} parameter variations in biological systems, and their effects on the preservation of biological functions and regimes.

In this work, to study the robustness of neurological systems with respect to parametric uncertainty, we resort to \textit{probabilistic} robustness analysis (PRA) \cite{tempo2013randomized}.
Differently from deterministic methods such as sensitivity and bifurcation analysis, and from classic parametric robustness analysis, PRA allows us to consider probabilistic parametric uncertainty in neurological models and more accurately analyse the effects of randomness in parameter realisations on the resulting (stochastic) behaviour of neural systems. In fact, PRA of a system whose uncertain parameters can vary, within a known bounding set in the parameter space, according to a known probability distribution (which encodes likelihood belief, estimation error, or population variability) has several advantageous properties.
\textbf{(i)} PRA identifies uncertainty regions in the parameter space where the same qualitative system behaviour is preserved \cite{goldman2001global, foster1993significance,beer1999evolution,marder2011variability} in probability. PRA can not only assess whether a system preserves a regime of interest in expectation in the presence of \textit{given} parametric uncertainties affecting its nominal parameters, but also \textit{quantify how much} parametric uncertainty affecting its nominal parameters a system can withstand, while still preserving a regime of interest in expectation.
\textbf{(ii)} PRA helps interpret the emergent \emph{average} effects of probabilistic parametric uncertainty: it allows to quantify, predict and analyse the \emph{mean} neural activity of ensembles of heterogeneous neurons, which is considered to be a more robust and reliable proxy for neural functions than the activity of single neurons, individually more fragile and prone to malfunctioning \cite{pouget2000information}.
Importantly, the signal associated with nominal deterministic values of the parameters may be completely different from the mean signal realisation obtained when the parameters are subject to parametric uncertainty within a parameter bounding set that contains the nominal parameter values (see Supplementary Figure~S4); hence, for instance, regime identification is much more difficult, compared to the deterministic setting.
If all parameter realisations in the bounding set yield the same qualitative behaviour (as in deterministic robustness analysis), then the mean realisation in the set is more likely, but is not guaranteed, to preserve the same qualitative behaviour.
These observations make the question of mean regime preservation under probabilistic parametric uncertainty much more intricate than its deterministic robustness counterpart.
In particular, studying the \textit{deterministic} realisation corresponding to the \textit{average} of multiple values of the system parameters cannot provide any meaningful information on the \textit{mean} system realisation. Indeed, for systems in neuroscience it is well known that the signal associated with the mean of multiple parameter values may not be representative of the regime associated with each of the original parameter values \cite{golowasch2002failure} (see, \textit{e.g.}, Supplementary Figure~S3). Systematic PRA tackles this complexity.
\textbf{(iii)} When comparing models to data, in-vivo recording methods such as electro-encephalograms (EEG) or local field potentials (LFP) measure the mean signals over batches of neurons \cite{perinelli2020chasing, lucas2024topological}. Therefore, PRA is a promising tool to understand whether regimes obtained from single-neuron models can robustly predict observed ensemble recordings, and to improve the synergy between mathematical models and real-life data through more reliable parameter fitting.
\textbf{(iv)} The ability to detect models that reproduce observed neurological phenomena only for specific parameter choices and cannot accommodate even reasonably small parameter variations while preserving the intended neurological behaviour, as opposed to models that display robust neurologically-meaningful regimes in the face of natural parametric uncertainty, makes PRA precious to falsify models and guide model selection, also in combination with suitably designed experiments \cite{goldman2001global}.

In this paper, we develop a novel methodology for the efficient and systematic PRA of uncertain neuronal systems and we demonstrate its efficacy on well-established dynamical models from mathematical neuroscience: the Hindmarsh-Rose (HR) model \cite{hindmarsh1984model} and the Jansen-Rit (JR) model \cite{jansen1995electroencephalogram}.
We consider a class of models of the form $\dot{\mathbf{x}}=f(\mathbf{x},Z)$ subject to probabilistic parametric uncertainty, where the parameters in $Z$ are independent random variables characterised by a known uniform probability distribution, supported on a bounded set that contains the nominal parameter values. Then, the state $\mathbf{x}(t;Z)$ is a stochastic processes and each realisation of the system trajectory depends on a realisation of the parameters in $Z$.
In this probabilistic framework, we aim to identify regions in the parameter space where the regime associated with the deterministic nominal model is preserved in expectation despite parametric uncertainties, so as to distinguish between regimes and quantify the largest uncertainty level compatible with probabilistic robustness of a regime of interest. 
To assess probabilistic robustness, we focus on time-averages (means) of output signals of interest (obtained from the imposed parameter distributions); to efficiently compute mean neural activities under parametric uncertainty, we resort to the generalised polynomial chaos (gPC) method \cite{xiu2003modeling}, which has two major advantages: first, it effectively incorporates parametric uncertainty into the analysed models and provides explicit analytic expressions that directly capture the uncertainty effects;
second, it provides a favourable trade-off between accuracy and computational costs \cite{sutulovic2024efficient}, often being much more efficient than popular Monte Carlo methods, a commonly used alternative for model-based PRA.
Then, we propose a computational methodology to extract patterns from the obtained mean time-series through the generation of recurrence plots \cite{eckmann1995recurrence}. The identification of distinct geometric patterns within recurrence plots enables to distinguish between regimes; we show that our proposed approach, based on \textit{blob counts}, successfully copes with the presence of parametric uncertainty (while other metrics often used for regime identification in deterministic models, such as inter-spike intervals, duty cycles or spike counts, fail to identify regimes and to infer reliable information about their preservation).
Finally, our methodology produces \textit{probabilistic regime preservation} (PRP) plots that contain essential information on the dynamical regimes induced by the system under study, and their robustness to parametric uncertainties, in a parameter space of interest. Differently from classical deterministic methods, such as bifurcation plots, the obtained PRP plots offer a probabilistic perspective on the mean behaviour of the system under investigation, and on the preservation of its regimes when the system parameters are affected by probabilistic uncertainty.
We present in detail our proposed computational methodology for PRA of neural systems in Section~\ref{sec:methodology}, after providing some necessary background on the considered models and on the gPC technique. Then, in Section~\ref{sec:results}, we apply our new approach to the HR and the JR models, and discuss the significance of the obtained PRP results for their robustness to parametric uncertainty. A concluding discussion is given in Section~\ref{Sec:ConcDiscuss}. Additional information and simulations are provided in the Supplementary Material.

\section{Preservation of regimes under parametric uncertainty}
\label{sec:methodology}

\begin{figure}[ht!]
	\centering
	\includegraphics[width=\linewidth]{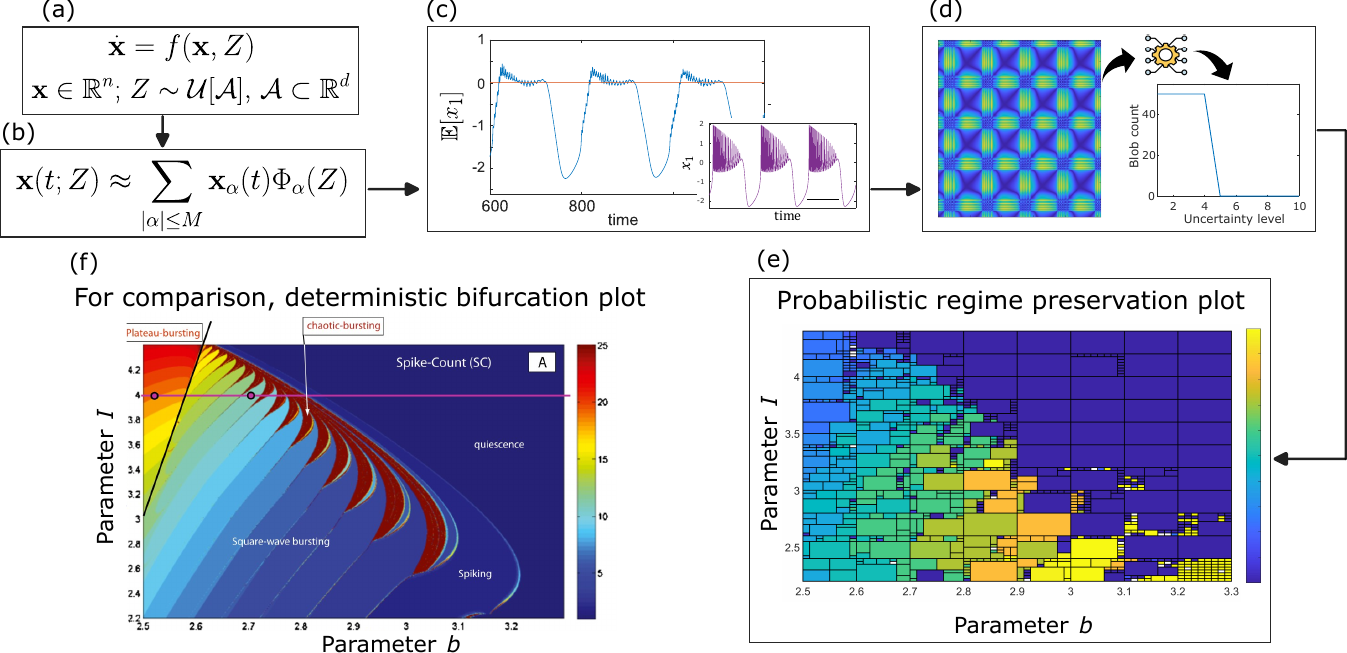}
	\caption{Workflow of our methodology for probabilistic robustness analysis (PRA) described in Section~\ref{sec:methodology}. (a) Consider a dynamical system of interest, where the evolution of the state vector $\mathbf{x} \in \mathbb{R}^n$ depends on a vector $Z$ of $d$ uncertain parameters, independent and distributed according to uniform distributions within known bounding intervals. (b) Derive a surrogate model of the system with the gPC method. (c) Based on the surrogate model, efficiently compute the mean system output (blue) with respect to the parametric uncertainty $Z$; the corresponding deterministic output of the nominal, uncertainty-free system is shown (purple) for comparison. (d) To unravel key geometric patterns within the time series associated with the mean system output, construct the corresponding recurrence plot (left). Then, automatically extract persistent features (blob count) of recurrence plots corresponding to increasing uncertainty levels, so as to assess regime preservation: the regime is preserved for all the uncertainty levels that achieve a similar blob count, while it is no longer preserved once the blob count drops (right). (e) Capture the overall information about dynamical regimes and their robustness to parametric uncertainty in a probabilistic regime preservation (PRP) plot, where the colour encodes blob counts and the length of each side of the rectangles quantifies how much uncertainty the system can withstand in the corresponding parameter value while maintaining the same regime in expectation. (f) For comparison, in the same parameter space as the PRP plot, we present a bifurcation plot (image courtesy of Springer, see \cite{barrio2011parameter} for additional details) that identifies the regimes resulting from different deterministic values of the parameters.
    }
	\label{fig:gPC_prob_rob_pipeline}
\end{figure}

This section describes our novel methodology for the PRA of systems in neuroscience subject to parametric uncertainties, leading to the calculation of PRP plots, as shown in Figure~\ref{fig:gPC_prob_rob_pipeline}.

\subsection{Class of considered systems subject to parametric uncertainties}\label{sec:models}

We consider models from mathematical neuroscience that are described by an ordinary differential equation (ODE) system of the form $\dot{\mathbf{x}}=f(\mathbf{x},Z)$, where $\mathbf{x}\in\mathbb{R}^n$ is the state vector and the uncertain system parameters of interest are stacked in a parametric uncertainty vector $Z\in\mathbb{R}^d$.
We assume that the components of $Z$ are independent random variables, characterised by a known probability distribution supported on a bounded set representing the parameter space of interest for the considered model. Many different distributions of $Z$ may be handled by our proposed approach \cite{xiu2003modeling}. Since the uniform distribution has been observed to be the worst-case distribution for PRA \cite{BARMISH!!}, and we can theoretically transform any continuous random variable into a uniformly distributed one through a quantile transformation \cite{kallenberg2002foundations}, in the absence of additional information on $Z$, we assume that each component $Z_i$, $i=1,\ldots,d$, is \textit{uniformly} distributed over an uncertainty interval containing the nominal value $Z_i^*$. Precisely, we therefore have $Z_i \sim \mathcal{U}([Z_{i,{\min}},Z_{i,{\max}}])$, with $Z_{i,{\min}}\leq Z_i^* \leq Z_{i,{\max}}$, $i=1,\ldots,d$.
The resulting ODE model subject to parametric uncertainty is
\begin{equation}
    \dot{\mathbf{x}}=f(\mathbf{x},Z),\qquad Z\sim \mathcal{U}(\mathcal{A}), \quad \mathcal{A}=[Z_{1,{\min}},Z_{1,{\max}}]\times\ldots\times [Z_{d,{\min}},Z_{d,{\max}}] \subset \mathbb{R}^d,
\label{eq:ode_param_uncert}
\end{equation}
where $f \colon \mathbb{R}^n\times\mathbb{R}^d\rightarrow\mathbb{R}^n$, the hyper-rectangle $\mathcal{A}$ is the uncertainty bounding set and $\mathbf{x}(t;Z)$ is a stochastic process (Figure~\ref{fig:gPC_prob_rob_pipeline}a); in this probabilistic setting, different system evolutions are possible depending on the specific realisation of the parameters in $Z$.
In \eqref{eq:ode_param_uncert}, stochasticity arises from probabilistic uncertainty in the system parameters, and not from a stochastic process driving the dynamics as it happens for instance in stochastic differential equations \cite{oksendal2013stochastic}.

Our approach can be applied to any system subject to parametric uncertainty, whose parameter space is not too high-dimensional, since the only bottleneck relates to computational cost.

To demonstrate our proposed methodology, we focus on two celebrated models from neuroscience: the Hindmarsh-Rose (HR) model for single-neuron dynamics \cite{hindmarsh1984model} and the neural mass mean-field Jansen-Rit model for a single cortical column \cite{jansen1995electroencephalogram}.

The HR model \cite{hindmarsh1984model} is widely used to reproduce the spiking and bursting dynamics of single neurons observed in experimental data \cite{gu2013biological}; however, its robustness to probabilistic parametric uncertainty has not been studied so far. In the adimensionalised ODEs of the HR model
\begin{equation}
    \begin{cases}
        \dot x_1(t) = x_2(t) - a x_1^3(t) + b x_1^2(t) - x_3(t) + I,\\
        \dot x_2(t) = c - d x_1^2(t) - x_2(t),\\
        \dot x_3(t) =  r[s(x_1(t) - x_R) - x_3(t)],
    \end{cases}
    \label{eq:HR}
\end{equation}
variable $x_1(t)$ represents the neuron membrane potential, $x_2(t)$ a fast recovery current, and $x_3(t)$ a slow adaptation current that enables bursting (\textit{i.e.}, trains of multiple membrane potential spikes). $I$ is an injected current, $x_R$ is the reference resting membrane potential, $r$ is the timescale ratio between fast and slow currents, while parameters $a$, $b$, $c$, $d$, $s$ are related to the system nullclines \cite{hindmarsh1984model}. We consider $Z_{\text{HR}} = [b,I]^\top$ as the parameters of interest, and set the other parameters to the standard values in \cite{hindmarsh1984model}: $a=1$, $c=1$, $d=5$, $s=4$, $x_R=-8/5$, $r=0.01$. In a \textit{deterministic} setting, for different values of the system parameters, the state trajectories $\mathbf{x}(t)=[x_1(t),x_2(t),x_3(t)]^\top$, $t\geq 0$, of this relatively simple model can exhibit multiple qualitatively distinct behaviours, which we call \textit{dynamical regimes}. We focus on the membrane potential, $y(t) = x_1(t)$, as an output signal of interest; examples of realisations of $x_1(t)$ for different parameter values that pertain to different dynamical regimes, ranging from quiescence to bursting, are shown in Figure~\ref{fig:deterministic_RP} (top).

The JR model \cite{jansen1995electroencephalogram} is a neural mass mean-field model that describes, rather than the dynamics of individual neurons, the generation of EEG signals by a single cortical column. In particular, the cortical column is represented through the interaction of the mean-field activity of three neuronal populations: pyramidal neurons, excitatory interneurons and inhibitory interneurons.
In the corresponding ODE system
\begin{equation}
    \begin{cases}
        \dot{x}_1 = x_4 , \qquad \dot{x}_4 = A a S(x_2 - x_3) - 2a x_4 - a^2 x_1 , \\
        \dot{x}_2 = x_5 , \qquad \dot{x}_5 = A a [p + C_2 S(C_1 x_1)] - 2a x_5 - a^2 x_2 , \\
        \dot{x}_3 = x_6 , \qquad \dot{x}_6 = B b C_4  S(C_3 x_1) - 2 b x_6 - b^2 x_3 ,
    \end{cases}
    \label{eq:JR}
\end{equation}
the states $x_i(t)$, $i=1,2,3$, are neural potentials ($x_1$ is the mean excitatory postsynaptic potential, $x_2$ is the mean membrane potential of pyramidal cells, $x_3$ is the mean inhibitory postsynaptic potential), while $S(V) = \nu_{M} / \left(1+e^{r(V_0 - V)}\right)$ is a sigmoidal function, where $\nu_{M}$ is the maximal firing rate of the family of neurons, $V_0$ is the excitability threshold of the population, $r$ is the firing threshold variance (\textit{i.e.}, the slope of the sigmoid at $V_0$). Parameters $A$ and $B$ are the maximum amplitude of excitatory and inhibitory postsynaptic potentials, respectively, while $a$ and $b$ are lumped parameters related to time constants in the dendritic network for excitatory and inhibitory synapses, respectively. All connectivity constants $C_i$, describing the strength of the links between pyramidal neurons and inhibitory and excitatory interneurons, are proportional to a single parameter $C$: $C_1 = C$, $C_2 = 0.8C$, $C_3 = 0.25C$, $C_4 = 0.25C$, with $C=135$ \cite{jansen1995electroencephalogram}.
We investigate $Z_{\text{JR}}=[A,B]$ as the parameters of interest, for varying values of $p$ and $C$, while all other parameters are fixed as in \cite{jansen1995electroencephalogram}: $a=100$, $b=50$, $V_0=6$, $\nu_M=5$, $r=0.56$. 
In a deterministic setting, the EEG-like output signal $y(t)=x_2(t)-x_3(t)$ exhibits different qualitative dynamical behaviours (regimes) as shown in Supplementary Figure~S2.

We consider systems \eqref{eq:HR} and \eqref{eq:JR} subject to probabilistic parametric uncertainties as in \eqref{eq:ode_param_uncert}. In particular, the HR model can be cast in the form \eqref{eq:ode_param_uncert} with $\mathbf{x}=[x_1(t),x_2(t),x_3(t)]^{\top}$, $Z=Z_{\text{HR}}=[b,I]^\top$, $f\left(\mathbf{x},Z_{\text{HR}}\right)$ given by the right-hand side of \eqref{eq:HR} and $\mathcal{A}$ being a rectangle that includes the nominal parameter values $Z_{\text{HR}}^*=[b^*,I^*]^\top$.
Similarly, for the JR model, $\mathbf{x}=[x_1(t),x_2(t),x_3(t),x_4(t),x_5(t),x_6(t)]^{\top}$, $Z=Z_{\text{JR}}=[A,B]^\top$, $f\left(\mathbf{x},Z_{\text{JR}}\right)$ is given by the right-hand side of \eqref{eq:JR} and $\mathcal{A}$ is a rectangle that includes the nominal parameter values $Z_{\text{JR}}^*=[A^*,B^*]^\top$.

To illustrate the next steps of our methodology, in this section we focus on the HR model \eqref{eq:HR}; the results for the JR model \eqref{eq:JR} are presented in Section~\ref{sec:results}.

We consider the HR parameter space $[b_{\min}, b_{\max}]\times[I_{\min},I_{\max}]=[2.5,3.3]\times[2.2,4.4]$. Figure~\ref{fig:gPC_prob_rob_pipeline}f shows the corresponding bifurcation diagram, from \cite{barrio2011parameter}, in the deterministic setting: regions with different colours correspond to deterministic parameter values that result in regimes with different characteristics (exemplified in Figure~\ref{fig:deterministic_RP}). In our numerical experiments, we choose nominal parameter values $[b^*,I^*]^\top$ that lie within the interior of a region associated with one of the deterministic nominal regimes in Figure~\ref{fig:gPC_prob_rob_pipeline}f. Probabilistic parametric uncertainty affecting the nominal parameter values may lead to loss of regime identifiability from the mean output trajectory realisation (see Figure~\ref{fig:gPC_prob_rob_pipeline}c and Supplementary Figure~S4 for examples).
Our goal is thus to develop a computational framework that allows us to detect regimes in the presence of parametric uncertainty and identify the highest possible uncertainty level (corresponding to the largest possible neighbourhood of the nominal parameters, given predefined specifications) where the nominal regime, exhibited by the system output under the \emph{deterministic} nominal parameters, is preserved in expectation. To this aim, we design dedicated metrics and apply them to the mean output signal, so as to relate it to the nominal regime.

\subsection{Generalised polynomial chaos (gPC) surrogate models}\label{sec:gPC}

To compute statistics of the stochastic process $\left\{\mathbf{x}\left(t;Z\right) \right\}_{t\geq 0}$ described by \eqref{eq:ode_param_uncert}, we employ a gPC approximation \cite{xiu2003modeling} to obtain a surrogate model of \eqref{eq:ode_param_uncert}, based on polynomial regression. The gPC surrogate model is simpler to integrate and retains the probabilistic information needed to efficiently calculate the mean process $\left\{\mathbb{E}\left[\mathbf{x}\left(t;Z\right) \right]=:\mathbb{E}[\mathbf{x}](t) \right\}_{t\geq 0}$, where $\mathbb{E}$ denotes the expectation with respect to $Z\sim \mathcal{U}(\mathcal{A})$.
This gPC-based approach has been recently shown to provide surrogate models characterised by a favourable trade-off between computational effort and accuracy, for relevant systems in neuroscience at different spatial scales \cite{sutulovic2024efficient}.

To introduce the gPC methodology, consider a general ODE system \eqref{eq:ode_param_uncert} subject to parametric uncertainty. The state trajectory $\left\{\mathbf{x}(t;Z)\right\}_{t\geq 0}$ is a stochastic process that depends on the parametric uncertainty $Z\in \mathbb{R}^d$, and we wish to efficiently compute the mean state trajectory $\mathbb{E}[\mathbf{x}] \colon \mathbb{R}_{\geq 0}\rightarrow\mathbb{R}^n$, where $\mathbb{E}[\mathbf{x}](t):=\frac{1}{\text{size}(\mathcal{A})}\int_{\mathcal{A}}\mathbf{x}(t;\xi)d\xi$ and $\text{size}(\mathcal{A})=\prod_{i=1}^d (Z_{i,{\max}}-Z_{i,{\min}})$.

We assume that the stochastic process $\left\{\mathbf{x}(t;Z)\right\}_{t\geq 0}$ has finite variance, which is not restrictive for mathematical models in neuroscience, since any meaningful model would take into account biophysical constraints, thus producing bounded state trajectories. Specifically, we assume that for each $t\geq 0$ (outside perhaps of a zero measure set), every component of $\mathbf{x}(t;\cdot)$, treated as a function of $Z$, belongs to the space $L_{\mathcal{U}}^2(\mathcal{A}) = \{\psi \colon \mathbb{E}[\psi^2]:=\frac{1}{\text{size}(\mathcal{A})}\int_A\psi^2(\xi)d\xi<+\infty\}$. This is guaranteed if the function $\mathbf{x}(t;\cdot)$ is measurable and bounded. Since the components of $Z$ are independent, a natural basis for $L_{\mathcal{U}}^2(\mathcal{A})$ is the set of multivariable polynomials $\{\Phi_\alpha\}_\alpha$ that satisfy the orthogonality relation $\mathbb{E}[\Phi_\alpha \Phi_\beta] = \frac{1}{\text{size}(\mathcal{A})}\int_A\Phi_\alpha(\xi)\Phi_\beta(\xi)d\xi = \delta_{\alpha \beta}$, where $\delta_{\alpha \beta}$ is the $d$-variate Kronecker delta and $\alpha = (\alpha_1,\ldots,\alpha_d)\in\mathbb{N}_0^d$ is a multi-index. 
The basis of orthogonal polynomials $\{\Phi_\alpha\}_{\alpha}$ can be computed via a Gram-Schmidt orthogonalisation process \cite{xiu2003modeling}.
As proven in \cite{cameron1947orthogonal}, the following series representation holds in $L^2_{\mathcal{U}}(\mathcal{A})$:
\begin{equation}
    \mathbf{x}(t;Z) = \sum_{\alpha \in \mathbb{N}_0^d}\mathbf{x}_\alpha(t)\Phi_\alpha(Z),
\label{eq:gPC_series}
\end{equation}
where the choice of the basis $\{\Phi_\alpha\}_\alpha$ depends on the probability distribution of the random variables in $Z$, according to the Wiener-Askey table \cite{xiu2002wiener,xiu2003modeling} that associates probability distributions with classes of orthogonal polynomials. In our case, since the random variables in $Z$ are uniformly distributed, $\{\Phi_\alpha\}_\alpha$ is composed of Legendre polynomials.
Importantly, the series representation \eqref{eq:gPC_series} neatly decouples the effect of the uncertain parameters $Z$, confined to the basis of orthogonal polynomials $\{\Phi_\alpha(Z)\}_\alpha$, and the deterministic time dependence, contained in the spectral coefficients $\{\mathbf{x}_\alpha(t)\}_\alpha$. This greatly simplifies the computation of statistics of the random process $\mathbf{x}(t;Z)$; for instance, mean and variance are given by
\begin{equation}
    \begin{cases}
    \mathbb{E}[\mathbf{x}](t) = \mathbf{x}_\mathbf{0}(t)\, ,\\
     \sigma_\mathbf{x}^2(t) = \sum_{|\mathbf{\alpha}| \neq 0} \mathbf{x}_\mathbf{\alpha}^2(t) \, .
    \end{cases}
\label{eq:mean_and_var_computation}
\end{equation}
To obtain a tractable surrogate model for practical implementation (Figure~\ref{fig:gPC_prob_rob_pipeline}b), we truncate the series \eqref{eq:gPC_series} to an expansion order $M$, namely, we only retain indices with $|\alpha|:=\sum_{i=1}^d \alpha_i \leq M$, thus approximating $\mathbf{x}(t;Z)$ with $\mathbf{x}_M(t;Z) = \sum_{|\alpha| \leq M}\mathbf{x}_\alpha(t)\Phi_\alpha(Z)$. 
Therefore, construction of the surrogate model only requires finding the spectral coefficients $\{\mathbf{x}_\alpha\}_{|\alpha|\leq M}$. For this, we adopt the non-intrusive \textit{collocation} approach, which consists of computing $S_\text{C}\geq \binom{M+d}{d}$ system trajectory realisations and then estimating the spectral coefficients indirectly, \textit{e.g.}, via a least-squares approximation \cite{blatman2011adaptive,sutulovic2024efficient}. 
Since theoretical convergence guarantees are available for static maps of random variables \cite{xiu2010numerical}, but not for solutions to general ODE systems, it is important to numerically check the suitability of the gPC-based approximation, by computing the coefficients $\{\mathbf{x}_\alpha(t)\}_{\left|\alpha\right|\leq M}$ with increasing values of $M$, until the decimal expansion of a fixed initial set of coefficients $\{\mathbf{x}_\alpha(t)\}_{\left|\alpha\right|\leq M_0}$, with $M_0< M$, stabilizes to a prescribed order of decimal digits. For computational purposes, we implement gPC collocation via the PoCET toolbox in \textsc{Matlab} \cite{Petzke2020_PoCET}.

\subsection{Mean state trajectories and output signals}\label{sec:meanoutput}

Given the parametric uncertainty $Z$, the gPC approach allows us to efficiently compute the mean system trajectory $\mathbb{E}[\mathbf{x}](t)$ as in \eqref{eq:mean_and_var_computation}. The mean output trajectory $\mathbb{E}[y](t)$, where $y \in \mathbb{R}$ is a linear combination of the state variables, can then be obtained from $\mathbb{E}[\mathbf{x}](t)$.
For the HR model, Figure~\ref{fig:gPC_prob_rob_pipeline}c shows the time evolution of the deterministic output $x_1(t;Z^*)$ (inset, purple) with nominal parameters $Z^*=[b^*,I^*]^\top=[2.44,4.2]^\top$ associated with the plateau bursting regime, along with the mean system output $\mathbb{E}[x_1](t)$ (blue) obtained when $I=I^*=4.2$ and $b\sim\mathcal{U}([2.4,2.48])$.

Sampling the mean system output yields a time-series $\{\mathbb{E}[y](t_k)\}_k$, formed by a collection of average values at time instants $\{t_k\}_k=\{k \cdot \Delta t\}_k$, $k \in \mathbb{N}$, where $\Delta t>0$ is a fixed inter-sampling time.
Then, we need to design criteria to determine whether, in the presence of the uncertainty $Z$, the sampled mean output signal $\{\mathbb{E}[y](t_k)\}_k$ preserves the qualitative features that characterise the nominal regime.
Assessing regime preservation in the presence of probabilistic parametric uncertainty is particularly challenging: we cannot resort to the standard metrics employed in the deterministic case for regime identification, as for mean signals they either cannot be computed or lose their interpretability (see Supplementary Section~S2).
In the deterministic case, \textit{i.e.}, for fixed parameter values, standard simple metrics such as inter-spike interval, duty cycle or spike count, often employed to extract quantitative information from neural signals \cite{barrio2011parameter, izhikevich2007dynamical, jirsa2014nature}, are easy to compute. However, with probabilistic uncertainty, the mean output signal is obtained by averaging different realisations of the output signal, whose spikes and bursts occur at different time instants or intervals. Hence, standard metrics are not properly defined for mean signals, where it is not even clear what constitutes a spike or a burst (see Supplementary Figure~S5).
Thresholding procedures further hamper the generic use of such standard metrics \cite{pavlov2000extracting}.
To distinguish between different regimes, an alternative approach may be to consider the mean output signal as a time-series and compute its power spectral density (or some wavelet transform) to search for distinctive frequency components \cite{rosso2006eeg, meisel2012scaling} (Supplementary Figure~S9). However, through numerical investigations partially reported in Supplementary Section~S2, we verified the current inefficacy of these and other existing methods used for regime identification in deterministic models, which become generally inapplicable in the presence of probabilistic parametric uncertainties.

In deterministic realisations of the HR model, the plateau-bursting regime (D) in Figure~\ref{fig:deterministic_RP} is characterised by variable $x_1$ exhibiting oscillations, initiated after it reaches the value $x_1 \approx 0$, which last with gradually damped amplitude for a finite time interval, after which $x_1 \approx 0$ again.
If such a distinctive characteristic is also present in the mean output signal for a given parametric uncertainty (as in Figure~\ref{fig:gPC_prob_rob_pipeline}c), then we may say that the regime is preserved in a probabilistic sense \cite[Section~III.C]{sutulovic2024efficient}. Unfortunately, cases in which the mean signal preserves such properties of the corresponding nominal regime are extremely rare. For the HR model, this happens only for plateau-bursting and not for other regimes; for other systems, it may never happen.

Therefore, for regime identification in a probabilistic uncertainty setting, other general techniques or metrics are needed, which we propose next.

\subsection{Persistence of geometric structures in recurrence patterns}
\label{sec:persistence}

We describe here our novel methodology that integrates recurrence plots and automated blob counting to study preservation of dynamical regimes in expectation under probabilistic parametric uncertainty.
We continue to illustrate our methodology on the HR model \eqref{eq:HR}, where we consider as output of interest $y(t)=x_1(t)$, the membrane potential, whose distinct qualitative behaviours give rise to different dynamical regimes such as those shown in Figure~\ref{fig:deterministic_RP}.

\subsubsection{Recurrence analysis}\label{sec:recurrence}

\begin{figure}[t!] 
	\centering
        \includegraphics[width=1\linewidth]{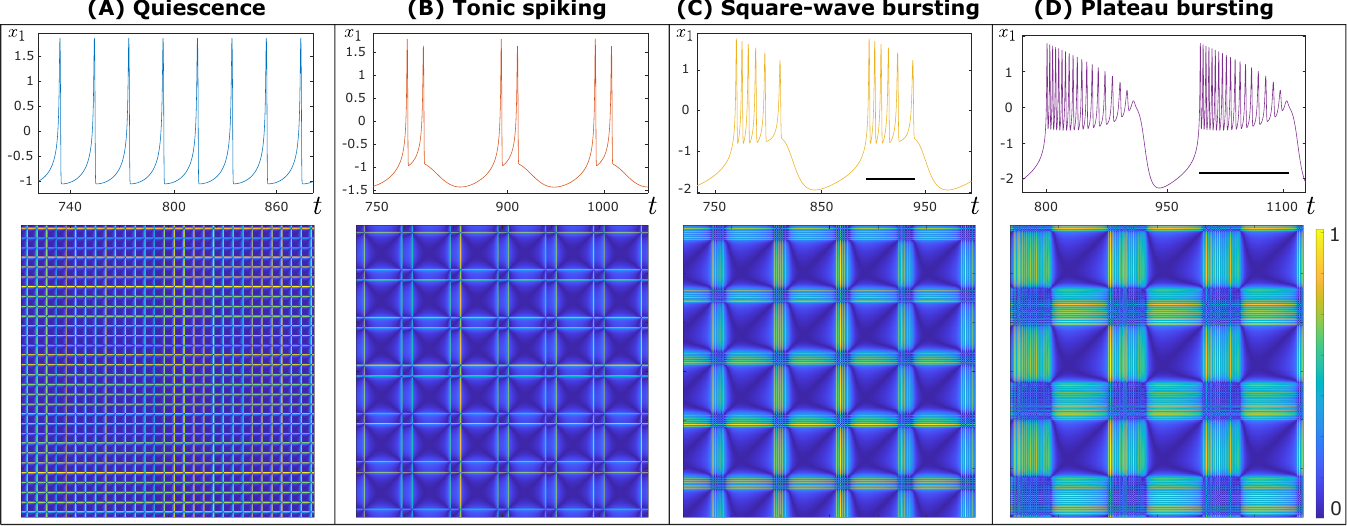}
	\caption{For the HR model \eqref{eq:HR}, deterministic realisations of $x_1(t)$ (top) and corresponding normalised recurrence plots in the time window $[600,1200]$ (bottom), with parameters $b \in [2.5,3.3]$ and $I \in [2.2,4.4]$ that yield different regimes: (A) $b=2.8$ and $I=4.2$, quiescence; (B) $b=3.1$ and $I=2.4$, tonic spiking; (C) $b=2.65$ and $I=2.4$, square-wave bursting; (D) $b=2.5$ and $I=4.0$, plateau bursting. In (C) and (D), the black horizontal line identifies a burst of spikes.}
	\label{fig:deterministic_RP}
\end{figure}

To extract information from mean output signals aimed at assessing regime preservation under parametric uncertainty, we resort to recurrence analysis \cite{lopes2021recurrence}.
An advantage of recurrence plots \cite{eckmann1995recurrence} lies in their ability to effectively visualise characteristic features of signals through constructed geometric patterns, which reveal qualitative signatures of, \textit{e.g.}, periodic and quasi-periodic behaviour, slowly evolving dynamics, stochastic-like fluctuations, or deterministic chaotic regimes \cite{marwan2007recurrence,prado2020maximum}.
To construct the recurrence plot, given $T$ samples of the mean output $\{\mathbb{E}[y](t_k)\}_{k=1}^T$ as our discrete-time signal of interest, we compute the distance matrix $D$ as
\begin{equation}
    D_{\ell h} = | \mathbb{E}[y](t_\ell) - \mathbb{E}[y](t_h) |, \quad 1\leq \ell,h \leq T.
\label{eq:rec_plot}
\end{equation} 
The entries of $D$ in \eqref{eq:rec_plot} yield an \textit{unthresholded} recurrence plot \cite[Subsection~3.2.5]{marwan2007recurrence}, which processes the distance matrix $D$ as an image with pixels coloured according to the values of its entries. Since the recurrence patterns are preserved under scaling, without loss of generality, $D$ is normalised by its maximum value, so that all of its entries $D_{\ell h}$ lie in $[0,1]$.
The recurrence plots for the HR model \eqref{eq:HR}, corresponding to distinct \emph{deterministic} realisations of $y(t)=x_1(t)$, the membrane potential, in different dynamical regimes, are shown in Figure~\ref{fig:deterministic_RP} (bottom); see also Figure~\ref{fig:gPC_prob_rob_pipeline}d, left. Recurrence plots allow us to visually distinguish between different dynamical regimes in a clear manner. This is also true for the JR model \eqref{eq:JR}, see Supplementary Figure~S2.

To study regime preservation, we analyse how the recurrence plot degrades when we increase the uncertainty level, \textit{i.e.}, the size of the support of the uniform distribution from which the uncertain parameters are randomly chosen. In particular, we consider a sequence of $N$ recurrence plots obtained for distributions with increasing uncertainty level: $\{Z\sim\mathcal{U}(\mathcal{A}_i)\}_{i=1}^N$, where the nominal value $Z^*\in \mathcal{A}_i$ for all $i=1,\ldots,N$ and $\mathcal{A}_{i+1}\supset \mathcal{A}_i$ for $i=1,\ldots,N-1$.

\begin{figure}[t!]
	\centering
        \includegraphics[width=1\linewidth]{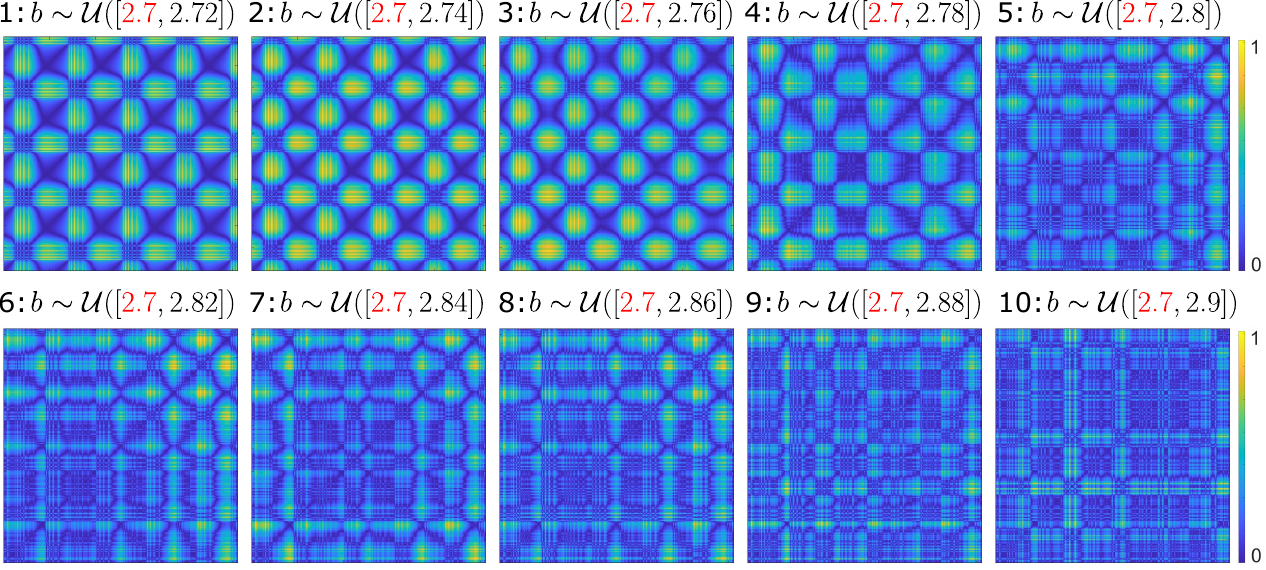}
	\caption{Degradation of recurrence plot patterns associated with the HR square-wave bursting regime in Figure~\ref{fig:deterministic_RP} (C) for increasing uncertainty levels $i=1,\ldots,N$, with $N=10$. Here $I=2.4$, while $b\sim\mathcal{U}([b^*,b^*+\Delta b_{i}])$ for the $i$-th uncertainty level, with $b^*=2.7$ (in red) and $\Delta b_i = \frac{i}{N} \Delta b_{\max}$ with $\Delta b_{\max}=0.2$ (i.e., $\Delta b_i=0.02i$ for $i=1,\ldots,10$). The nominal deterministic recurrence plot with $I=2.4$ and $b=b^*=2.7$ in shown in Supplementary Figure~S1.}
	\label{fig:RP_increasing_uncert}
\end{figure}

For the HR model \eqref{eq:HR}, we fix $I=2.4$ and the nominal $b^*=2.7$, consider uncertain $b$ with distributions $\{b\sim\mathcal{U}([b^*,b^*+\Delta b_{i}])\}_{i=1}^N$ having support of increasing size $\{\Delta b_{i}=\frac{i}{N}\Delta b_{\max}\}_{i=1}^N$ up to a maximum size $\Delta b_{\max}$, and assess whether the characteristic pattern associated with the nominal square-wave bursting regime persists in the recurrence plots for increasing $i=1,\ldots,N$, where the index $i$ encodes the uncertainty level.
Figure~\ref{fig:RP_increasing_uncert} shows the sequence of recurrence plots obtained with $\Delta b_{\max}=0.2$ and $N=10$.
The uncertainty pattern corresponding to the nominal square-wave bursting regime is preserved roughly up to the fourth uncertainty level, $b\sim \mathcal{U}([2.7,2.78])$. Then, the geometric patterns present in the nominal recurrence plot become significantly less prominent, and hence cannot be unequivocally associated with the square-wave bursting regime, at least based on visual inspection; the reason is that the uncertainty region spans a portion of the parameter space associated with trajectories having significantly different behaviours (\textit{e.g.}, with spikes occurring at vastly different time instants) and hence recurrence plots of the mean output signal $\mathbb{E}[x_1]$ do not retain a distinctive characteristic.

Recurrence analysis allow us to transform the problem of assessing regime preservation into image pattern recognition/retention. This task is often tackled using neural networks \cite{ripley2007pattern}, which however are not suitable in our case, because training and validation processes do not generalise well from one system to another, require extensive labelling, and are computationally heavy. A possible alternative would be unsupervised classification of images \cite{gadetsky2024let}, which however requires knowing the number of labels in advance; since we are not tackling the problem of \emph{global} classification of dynamical regimes in a parameter space, and we are only interested in a \textit{local} classification aimed at assessing local regime preservation under parametric uncertainty, we take a more direct and interpretable approach that is not based on machine learning algorithms, and leave their investigation for future research.

\subsubsection{Automated blob counting for recurrence plots}\label{Sec:BlobCountMethod}

Since recurrence plots exhibit a degradation of the geometric patterns associated with the nominal regime as the uncertainty level (\textit{i.e.}, the support of the uniform distribution) increases, the next step in PRA is to formulate a systematic, computable criterion to determine at which level of parametric uncertainty the geometric patterns in the recurrence plot are considered to be lost.

Let us denote by matrices $\left\{D_i \right\}_{i=1}^N$ as in \eqref{eq:rec_plot} the recurrence plots associated with the uncertainty intervals $\left\{\Delta b_i \right\}_{i=1}^N$, respectively.
Preliminary attempts for regime identification include computing a chosen matrix-related metric (\textit{e.g.}, maximum value, mean value, standard deviation and root-mean-square of all entries, spectral norm, or Frobenius norm) on each of the $D_i$'s and then attempting to associate the loss of the geometric patterns in the recurrence plot with a significant change in the metric (Supplementary Figures~S6 and S7).
Alternatively, by interpreting the $D_i$'s as matrices that describe probability mass functions defined on a common bounded domain, given two such matrices, one may compute the minimum “cost” to turn one into the other through Wasserstein distance, or mutual information quantifying their amount of shared information, or Jaccard index quantifying their degree of similarity (Supplementary Figure~S8).
For all of these approaches, among the mentioned metrics we found no definitive criterion that allowed us to reliably distinguish between different dynamical regimes (through distinct intervals of the metric values associated with distinct regimes) in the presence of parametric uncertainty, even for relatively small uncertainty levels, and reliably discern the loss of geometric patterns in recurrence plots (through a significant change in the metric) due to a significant increase in the uncertainty level, thereby identifying the uncertainty level for which the nominal regime is no longer preserved (see Supplementary Section~S2).

\begin{figure}[t!]
	\centering
	\includegraphics[width=1\linewidth]{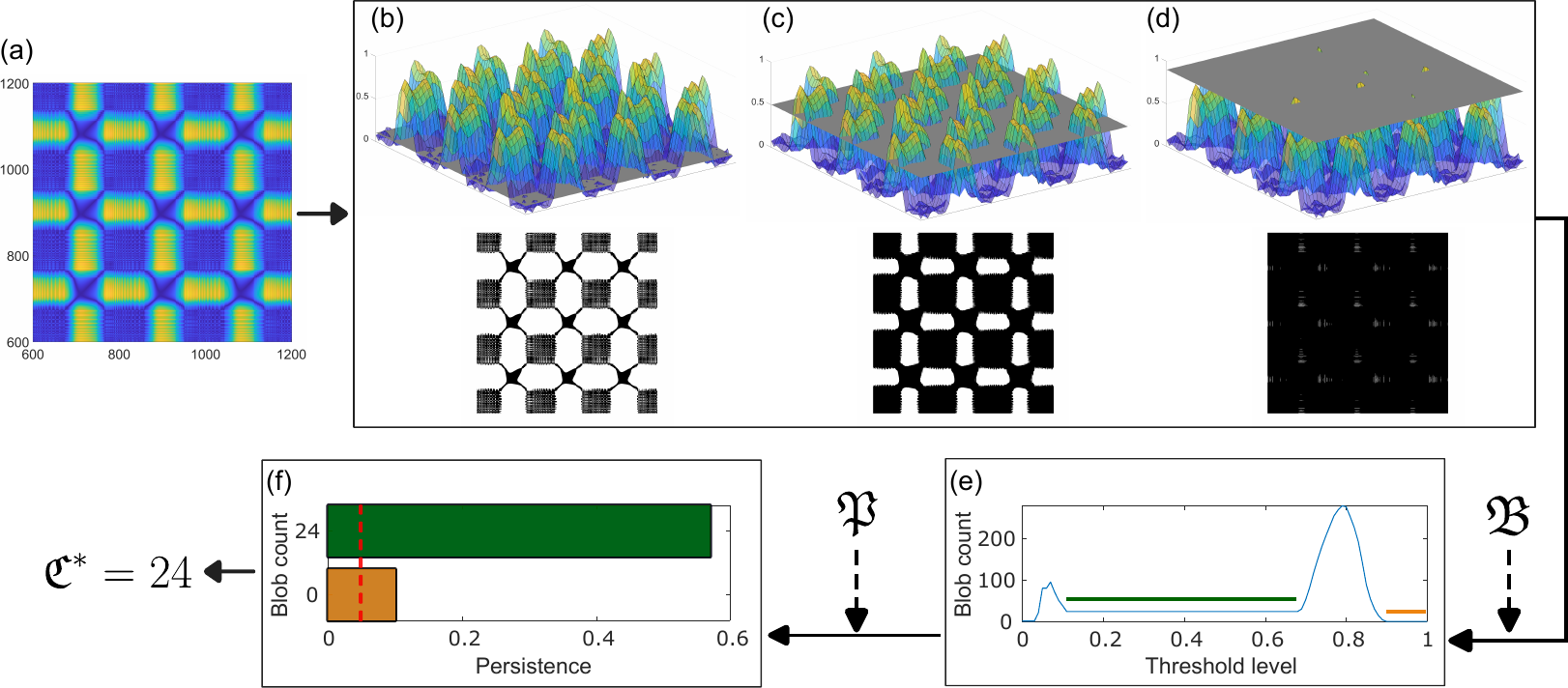}
	\caption{Automated blob count for (a) the recurrence plot $D$ of the mean output $\mathbb{E}[x_1](t)$ for the HR model \eqref{eq:HR}, with $b=2.5$ and $I\sim \mathcal{U}([3.6,3.8])$. (b)-(c)-(d) Matrix $D$ is visualised as a surface and different Boolean versions $\mathfrak{D}$ are obtained by processing with different thresholds: (b) $\vartheta=0.1$; (c) $\vartheta=0.5$; (d) $\vartheta=0.9$. (e) We discard blobs smaller than $\mathfrak{B}=150$ pixels and show the blob count $\mathfrak{C}$ as a function of $\vartheta$. (f) Persistence bar chart; among the blob counts with persistence larger than a user-defined minimum persistence $\mathfrak{P}=0.05$ (dashed red line), we select $\mathfrak{C}^*=24$ as it corresponds to lower thresholds; in fact, as shown in panel (e), $\mathfrak{C}=24$ is obtained for the thresholds highlighted in green, $\mathfrak{C}=0$ for those in orange.}
    \label{fig:blob_count}
\end{figure}

Therefore, we combine the derived recurrence plots with a novel quantitative metric for regime preservation/identification under probabilistic parametric uncertainty, based on persistence of the recurrence pattern of the mean output signal assessed through \textit{blob counts} (Figure~\ref{fig:gPC_prob_rob_pipeline}d, right).
Our new approach, described in Figure~\ref{fig:blob_count}, takes as an input a recurrence plot (Figure~\ref{fig:blob_count}a). The recurrence plot displays distinct yellow areas, which we call \textit{blobs}, that emerge from the background and are associated with the largest entries of $D$. The geometric patterns formed by such blobs distinctly characterise each given regime and we thus use them to assess regime preservation.
In particular, the blobs are fundamentally different in number between the different dynamical regimes, as can be observed in Figure~\ref{fig:deterministic_RP}, while their number is conserved for sufficiently small uncertainty levels, as shown in Figure~\ref{fig:RP_increasing_uncert}. Hence, the number of blobs is a reliable metric to assess the persistence of a regime, and distinguish between different regimes, in the presence of parametric uncertainty. 
To count the number of blobs in a recurrence plot with matrix $D \in [0,1]^{T \times T}$, we introduce a threshold $\vartheta \in (0,1)$ and we build the Boolean version $\mathfrak{D}$ of $D$, such that ${\mathfrak{D}}_{\ell h}=1$ if $D_{\ell h} \geq \vartheta$ and ${\mathfrak{D}}_{\ell h}=0$ otherwise. The resulting $\mathfrak{D}$ can be visualised as an image with black pixels associated with $0$ entries and white pixels associated with $1$ entries, as shown in Figure~\ref{fig:blob_count}b, c and d, where a three-dimensional visualisation of $D$ is also provided, with the threshold represented as a grey horizontal plane.

Given the Boolean matrix $\mathfrak{D}$, our proposed automated blob counting procedure counts the number of connected components of $\mathfrak{D}$ with value 1, corresponding to the white areas (see, \textit{e.g.}, Figure~\ref{fig:blob_count}b, c and d). As can be seen in Figure~\ref{fig:blob_count}b, where numerous small white dots are present, some of the connected components can be very small, and emerge due to numerical rounding errors and approximations. To avoid counting such small connected components, which may heavily skew the obtained blob count and its reliability, we introduce a user-defined minimum acceptable blob size $\mathfrak{B}$ (in terms of number of pixels), so that connected components of size smaller than $\mathfrak{B}$ are ignored in the counting process. The value of $\mathfrak{B}=150$ can be chosen as described in Supplementary Figure~S10. Therefore, the obtained blob count, denoted henceforth by $\mathfrak{C}$, corresponds to the number of connected components with size at least $\mathfrak{B}$.
As can be seen from Figure~\ref{fig:blob_count}b, c and d, different thresholds yield different blobs counts. Figure~\ref{fig:blob_count}e shows the blob count $\mathfrak{C}=\mathfrak{C}(\vartheta)$ as a function of the threshold $\vartheta$.
We proceed by selecting the most persistent values of $\mathfrak{C}$, 
namely, the blob counts that remain constant for the largest \textit{contiguous} interval of thresholds (denoted by the green and orange horizontal lines for the two most persistent values of $\mathfrak{C}$ in the example in Figure~\ref{fig:blob_count}e). We call \textit{persistence} of a blob count the length of the largest contiguous interval of thresholds for which the blob count is exhibited. For example, in Figure~\ref{fig:blob_count}e, the blob count of $24$ has persistence $0.58$, while the blob count of $0$ has persistence $0.1$. This is reflected, respectively, in the green and orange bars in Figure~\ref{fig:blob_count}f. We thus obtain a persistence bar chart for all persistent values of $\mathfrak{C}$.
To avoid considering blob counts with small persistence as representative of a dynamical regime, we further introduce a user-defined minimum acceptable persistence, $\mathfrak{P}$, which in our examples is set to be $\mathfrak{P} = 0.05$. Only blob counts with persistence higher than $\mathfrak{P}$ are considered as candidate blob counts for the considered recurrence plot. For example, in Figure~\ref{fig:blob_count}f both blob counts of $0$ and $24$ have persistence larger than $\mathfrak{P} = 0.05$.
Finally, if multiple blob count candidates have persistence larger than $\mathfrak{P}$, we select the blob count occurring at the lowest threshold as the actual blob count $\mathfrak{C}^*$. In fact, we posit that lower thresholds preserve more information of the original recurrence plot, because, as can be seen in Figure~\ref{fig:blob_count}d, higher thresholds lead to loss of the original geometric patterns, thereby resulting in a non-representative blob count. The same observation holds for low thresholds, which usually yield a single large connected component and, as a result, a non-informative blob count; therefore, we also discard the blob count $\mathfrak{C}=1$.
For example, in Figure~\ref{fig:blob_count}e and f, the blob count $\mathfrak{C}^*=24$, which is larger than $1$ and occurs at a lower threshold, is selected for the recurrence plot generated by the mean output signal of the HR model with $b=2.5$ and $I\sim \mathcal{U}([3.6,3.8])$.
If the blob counting procedure does not return any significant blob count value $\mathfrak{C}^*$, we do not assign any blob count to the corresponding recurrence plot, and consider that the regime has not been preserved (instead of a numerical value, we map it to a NaN; see Supplementary Figure~S11 for an example).

On a side note, the most suitable threshold $\vartheta$ to be used is not known a priori and likely depends on the specific dynamical regime.
As described in Figure~\ref{fig:blob_count}, we drew inspiration for threshold selection from \textit{persistent homology} \cite{chazal2021introduction}, which tracks how topological structures (such as connected components or holes) form and disappear, and employs such changes to distinguish between enduring patterns, as opposed to short-lived noisy structures, in data.
Similarly, we study how the blob count varies or persists through all possible thresholds $\vartheta \in (0,1)$. The most persistent blob count can be different from the most frequent blob count; thus, the employed persistence bar charts differ from simple histograms.

\subsubsection{Preservation of dynamical regimes}
\label{sec:preservation_limits}

Consider a system of the form \eqref{eq:ode_param_uncert} and nominal parameters $Z^*$.
Following the introduction of the blob counting methodology for recurrence plots in Section\ref{Sec:BlobCountMethod}, here we present a systematic method to determine the largest support $\mathcal{A}$ including $Z^*$ for the uniform distribution of $Z$ on which the nominal dynamical regime is preserved in expectation, according to the blob count metric.
In particular, to identify the largest uncertainty level (\textit{i.e.}, the largest size of $\mathcal{A}$) for which the associated recurrence plot retains a geometric pattern similar to that associated with the nominal regime, we can enlarge the size of the set $\mathcal{A}$, as shown in Figure~\ref{fig:RP_increasing_uncert}, as long as the resulting blob count $\mathfrak{C}^*$ is close, in an appropriate sense, to values of the blob count that characterise persistence of the nominal regime.
Hence, to assess regime preservation, we monitor persistence of the blob count as the uncertainty level is increased.

\begin{figure}[t!]
	\centering
        \includegraphics[width=0.9\linewidth]{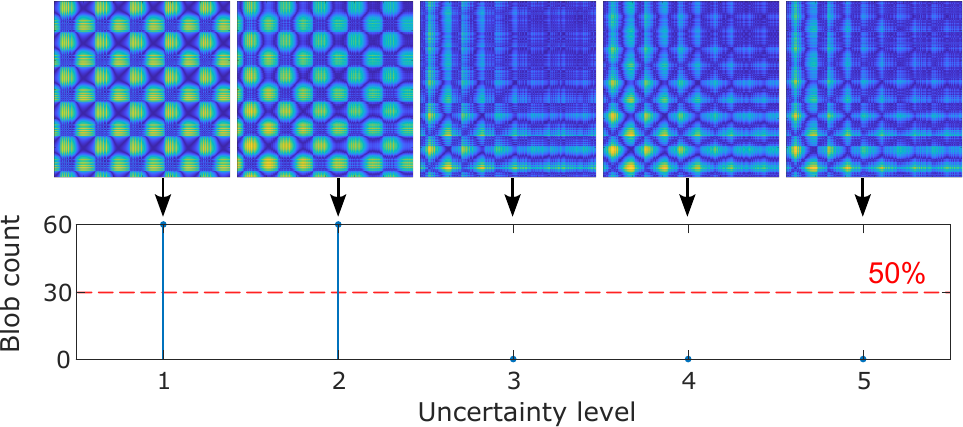}
	\caption{Blob counts with increasing uncertainty level for the recurrence plots of $\mathbb{E}[x_1]$ for the HR model \eqref{eq:HR}, with $I=2.8$ and $b\sim \mathcal{U}([2.7,2.7+0.03 i])$, $i=1,\ldots,5$. A significant blob count change occurs for $i=3$, when the blob count drops below $50$\% of the first blob count (dashed red line): the maximum tolerable uncertainty level is $i^*=2$.}
	\label{fig:blob_drop}
\end{figure}

Comparing the blob count obtained for the mean output signal at each uncertainty level with the blob count obtained for the deterministic output signal with nominal parameters $Z^*$ may be an unsuccessful approach: deterministic realisations and mean time-series can have completely different blob counts, even for small parametric uncertainties (see Supplementary Figure~S12), even though their recurrence characteristics are visually similar. This discrepancy is not surprising given the different nature of these signals and the frameworks in which they are employed, which offer distinct interpretations and perspectives (see Section~\ref{sec:introduction}).
Recall that the nominal parameter vector $Z^*$ is located in the \emph{interior} of a region in the parameter space that is associated with a precise dynamical regime, in the \emph{deterministic} setting. Therefore, in the presence of parametric uncertainty, we can safely assume that, if the parametric uncertainty at the first uncertainty level is small enough (\textit{i.e.}, $\mathcal{A}$ in \eqref{eq:ode_param_uncert} is sufficiently small), then the corresponding blob count is representative of the average dynamical regime around the realisation at $Z^*$. Denote by $\mathfrak{C}^*_1$ the blob count returned by the blob counting method in Section~\ref{Sec:BlobCountMethod} for the smallest uncertainty level. Similarly, let $\left\{ \mathfrak{C}^*_i\right\}_{i=2}^N$ be the blob counts associated with an increasing sequence of uncertainty levels, where the subscript $i$ grows with the uncertainty level. Let $\gamma \in (0,1)$ be a parameter chosen by the user. To detect loss of regime preservation, we look for the smallest index $i>1$ for which either $\mathfrak{C}^*_{i} < \gamma \mathfrak{C}^*_1$ or $\mathfrak{C}^*_{i} > (1+\gamma) \mathfrak{C}^*_1$.
Each of these inequalities is interpreted as a significant relative change in $\mathfrak{C}^*_{i}$ with respect to $\mathfrak{C}^*_1$. 
In general, $\gamma$ may be problem-dependent and is thus added as a user-defined parameter for the method.
In our examples, we use $\gamma = 0.5$ since, in most cases, the blob count drops to zero when the uncertainty level yields a significant degradation of the recurrence pattern (as in Figure~\ref{fig:blob_drop}); therefore, $\gamma=0.5$ is sufficient to exclude significant drops in blob count, while still allowing for small variations in the blob count as the uncertainty level increases. 
Supplementary Figure~S13 investigates the suitability of this criterion in the HR model.
Figure~\ref{fig:blob_drop} shows the sequence of blob counts, for increasing uncertainty levels, for the HR model \eqref{eq:HR} (fixing the left-end of the uncertainty intervals, as in Figure~\ref{fig:RP_increasing_uncert}).
The described procedure yields the maximum tolerable uncertainty level $i^*\leq N$ (\textit{i.e.}, the maximum $i\geq 1$ such that $\gamma \mathfrak{C}^*_1 \leq \mathfrak{C}^*_{i} \leq (1+\gamma) \mathfrak{C}^*_1$), corresponding to the maximum parametric uncertainty that the system can withstand while preserving the nominal regime, along with the maximum blob count $\mathfrak{C}^*_{\max}=\max_{i=1,\ldots,i^*}\mathfrak{C}^*_i$ associated with the preserved dynamical regime in the considered region of the parameter space.

\subsection{Construction of the probabilistic regime preservation plot}\label{sec:prp}

So far, we have proposed a method, based on blob counting in recurrence plots, that allows us to determine the persistence of a regime for system \eqref{eq:ode_param_uncert} with parametric uncertainty affecting the nominal parameters $Z^*$.
Consider now the case where multiple nominal parameter values $\{ Z^*_{(k)}\}_{k=1}^L$ are of interest in the parameter space $\mathcal{P}\subset\mathbb{R}^d$ that is relevant for the model under consideration, and assume that $\mathcal{P}$ is a hyper-rectangle. We can employ our proposed methodology to study regime preservation in the neighbourhood of each of the nominal parameter values $\{ Z^*_{(k)}\}_{k=1}^L$ in $\mathcal{P}$, thus assessing their local probabilistic robustness. The outcome of this analysis is a family of hyper-rectangles $\mathcal{A}_{(k)}\subset \mathcal{P}$, $k=1,\ldots,L$, where each $\mathcal{A}_{(k)}$ is the support of the maximum tolerable parametric uncertainty imposed on $Z^*_{(k)}$ in \eqref{eq:ode_param_uncert}, as defined in Section~\ref{sec:preservation_limits}, and hence the sets $\{\mathcal{A}_{(k)}\}_{k=1}^L$ indicate regions of probabilistic robustness under parametric uncertainty, according to our proposed methodology.
The resulting PRP plots for the parameter space $\mathcal{P}$ can be constructed, with two tailored approaches, in two different scenarios. In both cases, we can decide the number of uncertainty levels $N$ and the maximum size $\Delta Z_{\max}$ of the uncertainty interval for the parameters of interest, based on a trade-off between desired level of detail and sustainable computational effort.

\textbf{LE approach.} When no specific nominal parameter values of interest are given a priori, and we wish to gain information on probabilistic robustness properties of the regimes in the whole parameter space $\mathcal{P}$ of interest, we determine the set of $\{ Z^*_{(k)}\}_{k=1}^L$ in $\mathcal{P}$ with the aim of fully tiling $\mathcal{P}$ with hyper-rectangles $\{\mathcal{A}_{(k)} \}_{k=1}^L$.
In particular, we adopt a “left-end” (LE) approach that builds upon the same steps employed in the running example of the HR model \eqref{eq:HR} in Figures~\ref{fig:RP_increasing_uncert} and~\ref{fig:blob_drop}.
We set the nominal value $Z^*_{(1)} \in \mathcal{P}$ as the corner of the hyper-rectangle $\mathcal{P}$ with the smallest parameter values, and fix it as the “left-end” corner (with all entries being the smallest) of a sequence of hyper-rectangles that represent increasing parametric uncertainties. Then, we use the procedure described in Section~\ref{sec:preservation_limits} to find the blob count $\mathfrak{C}^*_{\max,(1)}$ and the maximum tolerable uncertainty level $i^*_{(1)}$ in order to determine $Z^*_{(2)}$ for the next iteration. We repeat the procedure until the whole $\mathcal{P}$ has been tiled with hyper-rectangles.
The LE approach yields a single PRP plot (see the examples in Figure~\ref{fig:gPC_prob_rob_pipeline}e and Figure~\ref{fig:HR_results}b1 and b2), in which the maximum blob counts $\{{\mathfrak{C}}^*_{\max,(k)}\}_{k=1}^L$ encoded by the colour of the hyper-rectangles $\{\mathcal{A}_{(k)}\}_{k=1}^L$ convey information about the associated dynamical regimes, while the size of the hyper-rectangles, each characterized by the corresponding level of uncertainty among $\{i^*_{(k)}\}_{k=1}^L$, represents how much parametric uncertainty can be imposed in the neighbourhoods of the nominal parameters $\{Z_{(k)}^*\}_{k=1}^L$ while guaranteeing preservation of the nominal regime in expectation: larger hyper-rectangles highlight regions in $\mathcal{P}$  that exhibit greater probabilistic robustness to parametric uncertainty.

The number of hyper-rectangles and their sizes are a priori unknown.
Still, the procedure is very easily implementable for a single uncertain parameter, namely if $\mathcal{P}\subset \mathbb{R}$, as it amounts to splitting a segment into smaller segments; examples are given in Supplementary Figure~S15. For $\mathcal{P}\subset\mathbb{R}^{n}$ with $n\geq 2$, however, the procedure is more complicated to iterate.
For $\mathcal{P}\subset\mathbb{R}^{2}$, which is the parameter space relevant for the HR model \eqref{eq:HR} that we use to present our methodology, the PRP plot is easy to visualise; see Figure~\ref{fig:HR_results}b1 and b2.
Details on the implementation of the LE approach to the HR model \eqref{eq:HR} can be found in Supplementary Figure~S14.

\textbf{FG approach.} When we are given multiple nominal parameter values of interest (for instance, associated with the design of experiments), we consider a grid of equidistant points $\{ Z^*_{(k)}\}_{k=1}^L$ in $\mathcal{P}$ that includes all the parameter choices of interest. If the chosen grid is “dense” enough, we may end up tiling the entire parameter space $\mathcal{P}$ of interest with hyper-rectangles $\{\mathcal{A}_{(k)}\}_{k=1}^L$.
We adopt a fixed-grid (FG) approach that considers a sequence of uncertainty intervals with increasing size, all having the nominal values $\{ Z^*_{(k)}\}_{k=1}^L$ \emph{at the centre}: for each component $j=1,\ldots,d$ of $Z_{(k)}$ we have
$\{Z_{(k),j}\sim\mathcal{U}([ Z_{(k),j}^*-0.5 (\Delta Z_{(k),j})_i,Z_{(k),j}^*+0.5 (\Delta Z_{(k),j})_i])\}_{i=1}^N$, where $(\Delta Z_{(k),j})_i$ is the size of the uncertainty interval for the $j$th component of the $k$th nominal parameter value at the $i$th uncertainty level.
The FG approach yields a pair of plots in the parameter space $\mathcal{P}$, as shown in the examples in Figure~\ref{fig:HR_results}c1 and c2 for the HR model: for each of the $\{Z_{(k)}^*\}_{k=1}^L$, one plot provides the corresponding blob count among $\{{\mathfrak{C}}^*_{\max,(k)}\}_{k=1}^L$, while the other plot provides the corresponding \textit{preservation percentage} $\{i^*_{(k)} \frac{100}{N}\}_{k=1}^L$, where $\{i^*_{(k)}\}_{k=1}^L$ are the  maximum tolerable uncertainty levels. Since we always execute at least one iteration, the preservation percentage is always at least $\frac{100}{N}$.

In both cases, the PRP plots capture changes in robustness for a dynamical regime without requiring a priori knowledge on the number and types of regimes exhibited by the system.

\section{Results}
\label{sec:results}

We present the results of our methodology for PRA applied to systems \eqref{eq:HR} and \eqref{eq:JR}, obtained with \textsc{Matlab} code run on a Windows 11 Dell Inspiron 16 laptop, with 16 GB RAM and 1.90 GHz Intel i5-1340P core.
We numerically integrate the systems using the function \texttt{ode45} with time step $10^{-2}$ for the HR model \eqref{eq:HR} and $10^{-4}$ for the JR model \eqref{eq:JR}, until final time $1200$ for HR and $2.5$ for JR, and discard an initial phase of $600$ time units for HR and $1.75$ time units for JR.

\subsection{Robustness of bursting at the single-neuron level}
We construct PRP plots for the single-neuron HR model \eqref{eq:HR} from \cite{hindmarsh1984model} and focus on the preservation of bursting activity in the membrane potential $x_1(t)$ under parametric uncertainty. Our interest is driven by the fact that bursting is a fundamental element of neural computations, to reliably transmit messages to post-synaptic neurons and overcome transmission failures \cite{lisman1997bursts}, and enable selective communication at the inter-neuron level \cite{izhikevich2003bursts}. Bursting is also important in triggering insulin secretion in $\beta$ cells within a pancreatic islet \cite{beauvois2006glucose} and release of growth hormone in cultured pituitary somatrophs \cite{tsaneva2007mechanism}, in sensory processing in the thalamocortical relay  \cite{sherman2002role}, and in ensuring temporal summation and long-term potentiation \cite{lisman1997bursts}. 
So, understanding how much parametric uncertainty the system can withstand while preserving bursting is relevant for many vital biological functions. To this end, we apply our methodology, described in Section~\ref{sec:methodology}, to $\mathbb{E}[x_1](t)$. Studying mean membrane potentials is justified by the principle of averaging \cite{van1992theory,faisal2008noise}, a natural mechanism in which the same noisy stimulus is processed in parallel by a redundant group of neurons, whose outputs are then averaged in the post-synaptic neurons, leading to a more fault-tolerant computation. Examples of averaging include neighbouring retinal ganglion cells processing visual signals \cite{puchalla2005redundancy} and stereocilia of auditory cells processing sounds \cite{kozlov2007coherent}.

In the HR model \eqref{eq:HR}, we consider three uncertain parameters: $I$, $b$ and $r$. Uncertainty in the input current $I$ arises from multiple parallel neurons receiving slightly different effective inputs due to heterogeneous synaptic drive. Variability in the parameters describing neuronal dynamics is due to inter-neuron heterogeneity. We focus on the parameters $b$, linked to the shape of the membrane potential nullcline, and $r$, a time ratio that is fundamental for bursting, and consider the parameter spaces $\mathcal{P}_1=[b_{\min}, b_{\max}]\times[I_{\min},I_{\max}]=[2.5,3.3]\times[2.2,4.4]$, with $r=0.01$, and $\mathcal{P}_2=[r_{\min}, r_{\max}]\times[I_{\min},I_{\max}]=[0.005,0.03]\times[2.4,3.4]$, with $b=3$. 

\begin{figure}[t!]
	\centering
        \includegraphics[width=1\linewidth]{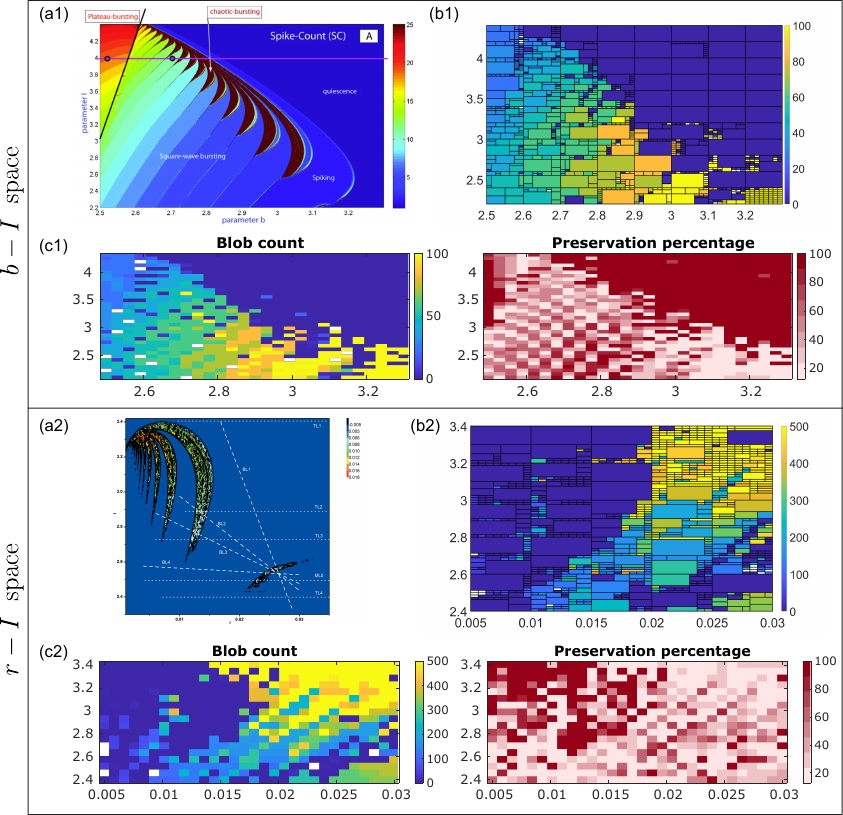}
	\caption{PRA of the bursting activity of membrane potential $x_1$ in the HR model \eqref{eq:HR} under probabilistic parametric uncertainty, in the parameter spaces $\mathcal{P}_1=[b_{\min}, b_{\max}]\times[I_{\min},I_{\max}]=[2.5,3.3]\times[2.2,4.4]$ with $r=0.01$ (top) and $\mathcal{P}_2=[r_{\min}, r_{\max}]\times[I_{\min},I_{\max}]=[0.005,0.03]\times[2.4,3.4]$ with $b=3$ (bottom). Mean signals are computed through gPC surrogate models via the collocation method, with maximum expansion order $M=5$ and $S_{\text{C}}=250$ collocation samples. In $\mathcal{P}_1$, $\Delta b_{\max}=0.1$ and $\Delta I_{\max}=0.2$; in $\mathcal{P}_2$, $\Delta r_{\max}=0.005$ and $\Delta I_{\max}=0.1$. In both cases, $N=8$ uncertainty levels are considered.
    In $\mathcal{P}_1$, we show:
    (a1) \textit{deterministic} dynamical regime classification based on bifurcation analysis, using spike count as a metric (image from \cite{barrio2011parameter}, courtesy of Springer); 
    (b1) PRP plot obtained with the LE approach; 
    (c1) PRP plot obtained with the FG approach, on a grid of $27\times36$ equidistant points.
    In $\mathcal{P}_2$, we show:
    (a2) \textit{deterministic} dynamical regime classification based on the largest Lyapunov exponent, which only highlights regions associated with the chaotic regime (image from \cite{gu2013biological}, courtesy of PLoS One);
    (b2) PRP plot obtained with the LE approach;
    (c2) PRP plot obtained with the FG approach, on a grid of $30\times18$ equidistant points.}
	\label{fig:HR_results}
\end{figure}

We show the PRP plots computed for $\mathcal{P}_1$ and $\mathcal{P}_2$ with the LE approach in Figure~\ref{fig:HR_results}b1 and b2, respectively, and with the FG approach in Figure~\ref{fig:HR_results}c1 and c2 respectively.
The plots in Figure~\ref{fig:HR_results}a1 \cite{barrio2011parameter} and a2 \cite{gu2013biological} display, in the same parameter spaces as the PRP plots, the regimes resulting from different fixed parameter values in a purely \textit{deterministic} setting, and hence offer an intrinsically different and complementary perspective.
The PRP plots visualise the preservation of dynamical regimes, identified through the blob count for the recurrence plots of the mean membrane potential $\mathbb{E}[x_1]$, in the \textit{probabilistic} setting of \eqref{eq:ode_param_uncert}. The blob count is encoded either in the colour of the rectangles in Figure~\ref{fig:HR_results}b1 and b2, or in the colour map in the left plots in Figure~\ref{fig:HR_results}c1 and c2.
The probabilistic robustness of nominal regimes in the presence of uniformly distributed parametric uncertainty is quantified either through the size of the rectangles in Figure~\ref{fig:HR_results}b1 and b2, or through the preservation percentage in the right plots in Figure~\ref{fig:HR_results}c1 and c2.

In the parameter space $\mathcal{P}_1$, the results stemming from our methodology reveal a “gradient” of colours associated with increasing blob counts, which can be observed both in Figure~\ref{fig:HR_results}b1 (LE approach) and Figure~\ref{fig:HR_results}c1, left (FG approach).
Although we do not know a priori which regime corresponds to each colour, comparison with the bifurcation plot in Figure~\ref{fig:HR_results}a1 \cite{barrio2011parameter}, which conveys information on the deterministic nominal regime associated with each point in the parameter space, allows us to identify the regime corresponding to a certain colour in the PRP plot: in Figure~\ref{fig:HR_results}b1 and Figure~\ref{fig:HR_results}c1, left, we can associate the upper-left light blue region with plateau bursting, the shades from turquoise to olive green to orange with square-wave bursting, and the yellow region with spiking, while the dark blue area encompasses both the chaotic and the quiescence regions, as discussed in depth later in this section.
The probabilistic robustness of each regime is quantified by the size of the rectangles in Figure~\ref{fig:HR_results}b1 (the longer the side of the rectangle, the more robust the system is with respect to the corresponding parameter) and by the preservation percentage plot in Figure~\ref{fig:HR_results}c1, right (the darker the colour, the higher the preservation percentage). In both cases, plateau bursting exhibits more robustness to parametric uncertainty than square-wave bursting, even though it is nominally emerging in a smaller portion of the parameter space $\mathcal{P}_1$. These findings are consistent with the robustness properties of bursting observed in a Hodgkin-Huxley-like model of pancreatic islets in \cite{smolen1993pancreatic}. 

In the parameter space $\mathcal{P}_2$, the deterministic plot based on Lyapunov exponents in Figure~\ref{fig:HR_results}a2 \cite{gu2013biological} highlights the coloured regions associated with a chaotic regime, which emerge from the blue background.
In our PRP plots, these regions have zero blob count (dark blue in Figure~\ref{fig:HR_results}b2 and in Figure~\ref{fig:HR_results}c2, left) and are particularly robust to parametric uncertainty (in Figure~\ref{fig:HR_results}b2, many dark blue rectangles have maximal area given by the user-defined choice; in Figure~\ref{fig:HR_results}c2, right, the regions have high preservation percentage).
However, a zero blob count also characterises the quiescence regime observed in the parameter space $\mathcal{P}_1$: also the quiescent dark blue areas in Figure~\ref{fig:HR_results}b1 and in Figure~\ref{fig:HR_results}c1, left, are very robust (large rectangles in Figure~\ref{fig:HR_results}b1 and high preservation percentage in Figure~\ref{fig:HR_results}c1, right).
In fact, discerning meaningful patterns from mean chaotic state trajectories in the presence of parametric uncertainty is particularly challenging: in a chaotic regime, the mean membrane potential computed in a probabilistic setting is indistinguishable from a zero signal affected by random noise, because it results from averaging deterministic realisations of firing patterns of the membrane potential that are very dissimilar, even for parameter values that are numerically close. This yields a rapid degradation of the recurrence patterns and hence a zero blob count, even for low uncertainty levels, thus making the mean signal in the chaotic regime indistinguishable from that in the quiescence regime, as is apparent from Figure~\ref{fig:HR_results}b1, where no discontinuity is visible between the chaotic bursting and the quiescence regions (for another example, see Supplementary Figure~S4).
On the other hand, chaos disrupts the probabilistic robustness of the bursting regimes even for small uncertainty levels, as shown by the fragmented tiling on the diagonal of Figure~\ref{fig:HR_results}b1, on the border between square-wave bursting and chaotic bursting regions (see also Supplementary Figure~S16).
Our PRP plots also reveal other, non-chaotic regions in $\mathcal{P}_2$, corresponding to large blob counts, where however the associated regime has a low level of robust preservation under uncertainty.

\subsection{Robustness at the cortical column level}

\begin{figure}[t!]
	\centering
        \includegraphics[width=1\linewidth]{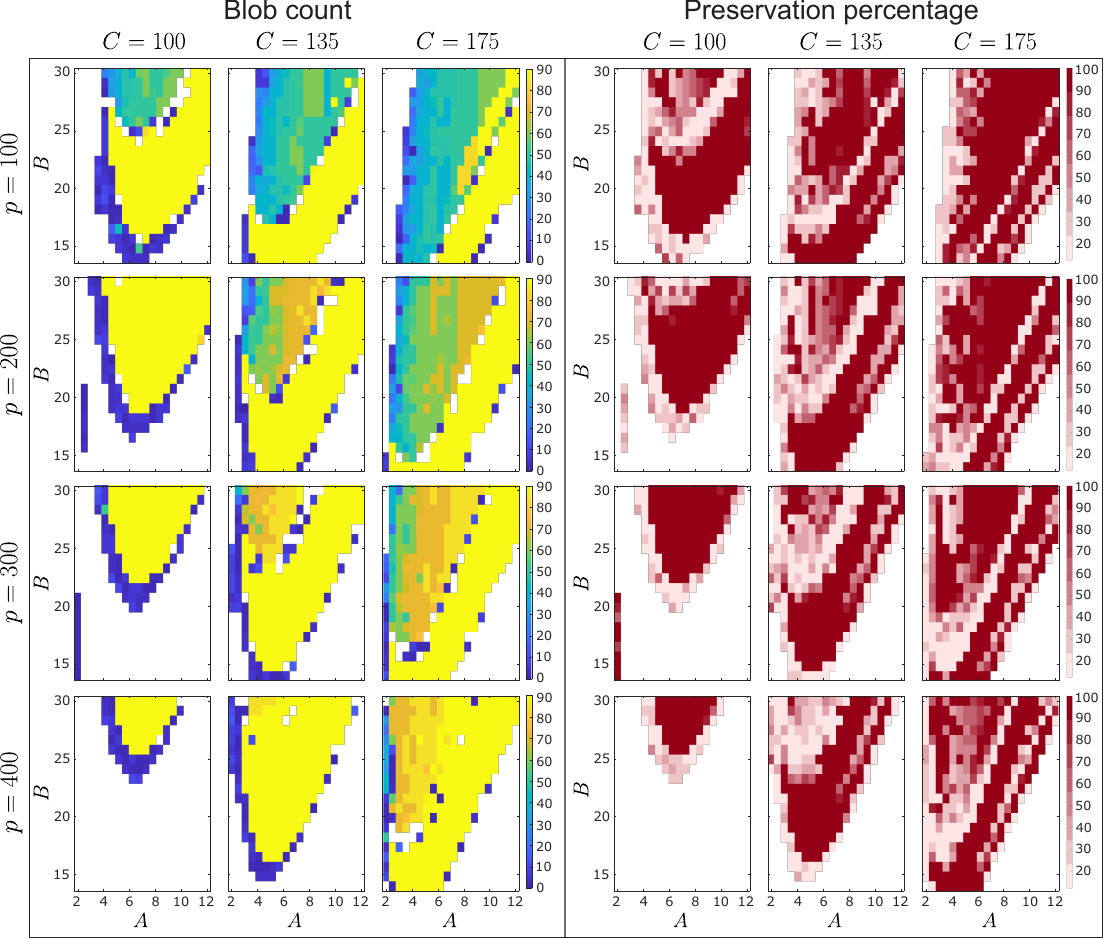}
	\caption{PRA of the EEG-like signal $x_2-x_3$ in the JR model \eqref{eq:JR} under probabilistic parametric uncertainty, in the parameter space $\mathcal{P}_{JR}=[A_{\min}, A_{\max}]\times[B_{\min},B_{\max}]=[2,12]\times[15,31]$, with $N=8$ uncertainty levels, $\Delta A_{\max}=2$ and $\Delta B_{\max}=2$, for different values of the input $p$ and the connectivity constant $C$. Mean signals are computed through gPC surrogate models via the collocation method, with maximum expansion order $M=5$ and $S_{\text{C}}=250$ collocation samples. PRP plots are obtained with the FG approach, on a grid of $20\times 20$ equidistant points.} 
	\label{fig:JR_A-B_space}
\end{figure}

The JR model \eqref{eq:JR}, along with its extensions including interneurons and multiple cortical columns to better reproduce the activity of different brain regions, has been thoroughly analysed via bifurcation analysis, to investigate average inhibitory synaptic gain and time constant of mean postsynaptic potentials \cite{subramaniyam2024sensitivity}, as well as influence of input strength and dendritic time constant in a multi-input model \cite{spiegler2010bifurcation}, and input strength and interconnection gains in two coupled models \cite{ahmadizadeh2018bifurcation}.
None of these works, however, considers parametric uncertainty, which is our focus.

We study the robustness of the EEG-like signal $y(t)=x_2(t)-x_3(t)$ with respect to parametric uncertainty in the amplitude of excitatory and inhibitory postsynaptic potentials, $A$ and $B$, for the single-cortical-column JR model \eqref{eq:JR} from \cite{jansen1995electroencephalogram}.
Figure~\ref{fig:JR_A-B_space} shows the resulting PRP plots constructed, with the FG approach, in the parameter space $\mathcal{P}_{JR}=[A_{\min}, A_{\max}]\times[B_{\min},B_{\max}]=[2,12]\times[15,31]$ also considered in \cite[Section~3.1]{jansen1995electroencephalogram}, for different values of the input strength $p$ and of the connectivity constant $C$.
Supplementary Figure~S18 provides additional PRP plots for other values of $C$, while PRP plots obtained with the LE approach are in Supplementary Figure~S17. Moreover, Supplementary Figure~S2 demonstrates the three characteristic regimes of the JR model \eqref{eq:JR}: alpha-wave (highest blob count, yellow regions in Figure~\ref{fig:JR_A-B_space}, left), low-frequency wave (intermediate blob count, turquoise to green to orange regions in Figure~\ref{fig:JR_A-B_space}, left), and no-wave (zero blob count in theory, white regions in Figure~\ref{fig:JR_A-B_space}, left).

The PRP plots in Figure~\ref{fig:JR_A-B_space} show that the alpha-wave regime is very robust to parametric uncertainty across different values of $p$ and $C$, as it is mostly associated with high preservation percentages. Regions associated with the low-frequency wave regime enlarge for increasing values of $C$, and are on average less robust to parametric uncertainty. The no-wave regime is characterised by a constant output signal; therefore, if the considered uncertainty set is entirely contained in a no-wave region, the mean neural activity yields in principle a zero blob count (persistent absence of recurrence pattern) that is robust to parametric uncertainty. In practice, however, some blobs may still be identified in the corresponding recurrence plot due to numerical fluctuations and due to normalisation of the distance matrix entries in \eqref{eq:rec_plot} so that they take values in $[0,1]$ (see Supplementary Figure~S2); however, this would result in a probabilistic robustness assessment that is not meaningful.
To tackle this problem, whenever at the first uncertainty level the difference between the maximum and the minimum values of $\mathbb{E}[y](t)$ is less than a pre-defined tolerance of $10^{-7}$, we assign a special value, encoded by white regions and associated with the constant no-wave regime, to both the blob count and the preservation percentage.

As mentioned earlier, zero blob counts (encoded in blue) can also arise from averaging very different deterministic realisations, which can lead to a mean signal that is indistinguishable from “random noise”; also in these cases, the condition $\max_t\{\mathbb{E}[y](t)\}-\min_t\{\mathbb{E}[y](t)\} < 10^{-7}$ may be satisfied if the “noise” amplitude is small enough, leading to a white dot instead of a blue one.
In fact, we observe that, in most PRP plots, a strip of blue dots corresponding to zero blob counts, and sometimes of white dots obtained through thresholding, clearly separates the regions associated with the alpha-wave and the low-frequency wave regimes, and also separates white regions associated with the no-wave regime from the regions associated with the other regimes; this happens because the corresponding uncertainty sets contain parameters that yield conflicting deterministic realisations, which do not allow for persistent recurrence patterns in the mean neural activity.

\section{Conclusions}\label{Sec:ConcDiscuss}

Focusing on dynamical systems in neuroscience, we have introduced a novel methodology for PRA aimed at quantifying how much probabilistic parametric uncertainty can be imposed on the nominal system parameter values, while ensuring that the mean system output preserves the qualitative recurrence pattern associated with the nominal dynamical regime.
Assessing how much uncertainty the system can withstand while still preserving the nominal regime in expectation is crucial to analyse interpretable mathematical models in neuroscience under parametric uncertainty and identify the allowable uncertainty ranges for which a life-sustaining regime is preserved.
We have employed the gPC approach to efficiently calculate mean neural signals through a simpler polynomial surrogate model.
The gPC method has been recently employed to generate surrogate models for uncertain systems in neuroscience \cite{ghori2023uncertainty,signorelli2024uncertainty,sutulovic2024efficient} and systems biology \cite{paun2025comparison,sumser2025exploiting,paulson2019fast,hu2018generalized}, and used for global sensitivity analysis of quantities of interest in \emph{specific} mathematical models or for parameter estimation from data. Differently from other works, here we have relied on gPC surrogate models as a key step in our proposed \emph{general} framework for the probabilistic robustness analysis of models in neuroscience.
To detect loss of regime preservation from the gPC-computed mean neural time-series, we have developed an algorithm based on the identification and the persistence of geometric patterns in the associated recurrence plots; this has been necessary to overcome the impossibility of adopting traditional neuroscientific metrics used for deterministic signals, which cannot be computed or lose their interpretability in a probabilistic setting.
From a computational perspective, to systematically identify geometric patterns, we have resorted to an automated count of the number of blobs, \textit{i.e.}, distinct regions with high values in recurrence plots; persistence of blob counts has been used as a metric for persistence of geometric patterns, and hence robust regime preservation.
Finally, we have proposed two different approaches to investigate probabilistic robustness to parametric uncertainty within a whole parameter space of interest through probabilistic regime preservation (PRP) plots, which convey information both on the dynamical regimes that are preserved and on the level of tolerable parametric uncertainty affecting various nominal parameter values.

In view of its generality and flexibility, our proposed methodology is widely applicable to gain deeper insight into uncertain dynamics in neuroscience, and beyond; we have showcased its application to two fundamental models of neural dynamics.

The PRP plots for the membrane potential in the single-neuron HR model \eqref{eq:HR} reveal that plateau bursting is consistently more robust to probabilistic parametric uncertainty than square-wave bursting.
Our novel methodology also provides an interesting insight: preservation of recurrence characteristics for mean state trajectories is unlikely in chaotic regions of a parameter space, as averaging completely different signals (arising even from very similar parameter values) leads to loss of any distinct geometric signature.
The work \cite{perinelli2020chasing} focuses on detection of chaos in human EEG brain signals and finds no evidence of chaotic behaviour, and our results (Figure~\ref{fig:HR_results}) suggest an explanation for this absence of evidence: since a single electrode measures the mean activity of several neurons, a chaotic regime would be very difficult to observe experimentally, even when some individual neurons exhibit a chaotic firing pattern as in \cite{gu2013biological}.

For the EEG-like signal generated by the single-cortical-column JR model \eqref{eq:JR}, we investigate the effect of uncertainty in the amplitude of excitatory and inhibitory postsynaptic potentials. PRP plots (Figure~\ref{fig:JR_A-B_space}) show how the regimes and the associated probabilistic robustness properties change for different values of the input and of the connectivity constant of the cortical column elements, and provide precious insight into the choice of the most appropriate parameter regions to consider when modelling the behaviour of cortical columns in a brain area of interest, based not only on the expected dynamical regimes, as in classical bifurcation analysis and in \cite{jansen1995electroencephalogram}, but also on the observed robustness properties under parametric uncertainties, unavoidable in experimental settings.

In addition to what we have shown in this work, PRP plots can be precious to identify the most sensitive direction in which the nominal values of uncertain parameters can vary so as to transition from a dynamical regime to another, similarly to the analysis in \cite{goldman2001global}. The novel information provided by our methodology is an estimate of the robustness to parametric uncertainty of the regime to be reached, which is fundamental for the design of control actions that safely steer the neural activity towards a desired functioning. For instance, to safely design an anaesthetic action that drives brain regions to a more relaxed dynamical regime \cite{kuhlmann2016neural}, we would like to minimise the changes with respect to nominal system parameter values, to avoid altering normal brain activity, while being reasonably sure to reach an anaesthetised state that is robust to parametric uncertainty, so that the control action can be effective under perturbations.

Future research directions include refinements of the proposed PRA framework that enrich the information content of PRP plots: higher-order moments of state trajectories can be easily retrieved with gPC \cite{lefebvre2020moment} (see, \textit{e.g.}, \eqref{eq:mean_and_var_computation}, where a formula for the variance of the stochastic process in \eqref{eq:ode_param_uncert} is also provided) and may be used to gain additional information that would, for instance, help distinguish between chaotic and quiescence regimes in a parameter space.

\section*{Code availability and Supplementary Material}
The code to reproduce all our results as well as the Supplementary Material can be found at \\\texttt{https://github.com/Uros-S/gPC-neuro-prob-robustness}

\section*{Funding}
Work supported by the European Union through the ERC INSPIRE grant (project number 101076926). Views and opinions expressed are however those of the authors only and do not necessarily reflect those of the European Union or the European Research Council Executive Agency. Neither the European Union nor the European Research Council Executive Agency can be held responsible for them. R. Katz also acknowledges support from the Alon Fellowship, awarded by the Council of Higher Education of Israel.

\section*{Declaration of competing interest}
The authors declare no competing interests.

 \bibliographystyle{elsarticle-num} 
 \bibliography{bibliography}






\end{document}